\documentstyle[psfig]{mn}

\title[Orbital periods of the binary sdB stars PG\,0940+068 and PG\,1247+554]
{Orbital periods of the binary sdB stars PG\,0940+068 and PG\,1247+554}
\author[P. F. L. Maxted et~al.]
       {P. F. L. Maxted$^1$, C. K. J. Moran$^1$, T. R. Marsh$^1$,
        and A.~A.~Gatti$^2$  \\
        $^1$University of Southampton, Department of Physics \& Astronomy,
        Highfield, Southampton, S017 1BJ, UK\\
        $^2$Imperial College of Science Technology and Medicine, 
         Blackett Laboratory, Prince Consort Road, London SW7 2BZ }
\date{Accepted 1999 
      Received 1999 }

\pagerange{\pageref{firstpage}--\pageref{lastpage}}
\pubyear{1999}

\newcommand{\Msolar}{\mbox{${\rm M}_{\odot}$}}
\begin{document}

\maketitle

\label{firstpage}

\begin{abstract}
 We have used the radial velocity  variations of two sdB stars previously
reported to be binaries to establish their orbital periods. They are
PG\,0940+068, (P=8.33d) and PG\,1247+554 (P=0.599d). The minimum masses of the
unseen companions, assuming a mass of 0.5\Msolar\ for the sdB stars, are
$0.090\pm0.003$\Msolar\ for PG\,1247+554 and $0.63\pm0.02$\Msolar\ for
PG\,0940+068. The nature of the companions is not constrained further by our
data. \end{abstract} 

\section{Introduction}
 Sub-dwarf B (sdB) stars are thought to be core helium burning stars with
core masses of about 0.5\Msolar\ and extremely thin ($\la0.02\Msolar$) hydrogen
envelopes (Heber 1986; Saffer et~al. 1994).  It is thought that the eventual
fate of an sdB star is to cool to form a white dwarf with a mass of about
0.5\Msolar, which is low compared to the typical mass for white dwarfs
(Bergeron, Saffer \& Liebert 1992). The formation of low mass white dwarfs,
and by implication sdB stars, is thought to involve a common envelope phase,
in which a companion to a red giant star in engulfed by the expanding outer
layers. The resulting friction results in the companion spiraling in towards
the core of the red giant, ejecting the envelope at the expense of orbital
binding energy (Iben \& Livio 1993). This provides a natural mechanism for
the loss of the surface hydrogen layers in sdB stars. 

 The binary fraction of sdB stars is expected to be high given the scenario
outlined above. Allard et~al. (1994) found that 31 of their sample of 100 sdB
stars show flat spectral energy distributions which indicate the presence
of companions with spectral types in the range late-G and early-M. They
infer a binary fraction for main-sequence companions of 54\,--\,66\,percent,
although the companions in their survey are over-luminous compared to normal
main-sequence stars. A similar conclusion was reached by Jeffery \& Pollacco
(1998) based on the detection of spectral features due to cool companions.
Companions to sdB stars can also be detected in eclipsing systems such as
HW~Vir (Wood \& Saffer 1999), PG\,1336-018 (Kilkenny et~al. 1998) and the
sdB\,--\,white dwarf binary KPD\,0422+5421 (Koen et~al. 1998). Saffer, Livio
\& Yungelson (1998) found that at least 7 of their sample of 46 sdB stars show
radial velocity variations. Several of these binaries have subsequently had
their orbital periods determined (Moran et~al. 1999), although further
observations are required to determine the nature of the companions in these
binaries.  

 In this paper we present two new orbital period determinations from radial
velocity measurements for binary sdB stars from the survey of Saffer, Livio \&
Yungelson (1998).

\section{Observations and reductions}
 Most of the data for this study comes from observations obtained with the 
2.5m Isaac Newton Telescope (INT) on the Island of La Palma. Spectra were
obtained with the intermediate dispersion spectrograph (IDS) using the 500mm
camera, a 1200 line/mm grating and a TEK CCD (charge coupled device) as a
detector. 

 Additional spectra of  PG\,0940+068 were obtained using three other
instruments.  Ten spectra were obtained with the  grating spectrograph on 1.9m
telescope at the South African Astronomical Observatory (SAAO). Grating No. 5
(1200 lines/mm) and a 1.2arcsec slit were used with the SIT1 CCD detector.
Four spectra were obtained with RGO spectrograph and 82cm camera on the
Anglo-Australian telescope (AAT). Full details of these spectra can be found
in Maxted \& Marsh (1999). 


Two additional spectra of PG\,1247+554 were obtained with the dual beam
spectrograph ISIS on the 4.2m William Herschel Telescope (WHT) on the
Island of La Palma. We used a 1200l/mm grating on the red arm to obtain
spectra of H$\alpha$ and HeI\,6678. The detector was a TEK CCD and the
slit width was 1.2arcsec.

For all these instruments and detectors, sensitivity variations were removed
using observations of a tungsten calibration lamp. Bias images show no signs
of any structure in the region of the spectrum so a constant bias level
determined from a clipped-mean value in the over-scan region was subtracted
from all the images. The exposure times used varied from 300s to 1200s.

\begin{table}
\caption{\label{ObsTable} Summary of the spectrograph/telescope combinations
used to obtain spectra for this study. The slit width used in each case is
approximately 1arcsec.}
\begin{tabular}{lllrr}
Date &Telescope & Spectro-&   Resolution & Sampling \\
 & & graph  & (\AA) & (\AA) \\
Jun 97 & INT & IDS &  1.0  & 0.4 \\
Feb 98 & INT & IDS &  1.0  & 0.4 \\
Mar 98 & SAAO 1.9m & --  & 0.8  & 0.4 \\
May 98 & AAT & UCLES & 0.10 & 0.04 \\
Jun 98 & AAT & RGO &  0.7 & 0.3  \\
Feb 99 & INT & IDS &  1.0  & 0.4 \\
Jul 99 & WHT & ISIS & 0.73 & 0.4  \\  
\end{tabular}
\end{table}

Extraction of the spectra from the images was performed automatically using
optimal extraction to maximize the signal-to-noise of the resulting spectra
(Horne 1986). Every spectrum was bracketed by observations of the internal
arc lamp at the position of the star. The arcs associated with each stellar
spectrum were extracted using the profile determined for the stellar image to
avoid possible systematic errors due to tilted spectra. The wavelength scale
was determined from a polynomial fit to measured arc-line positions.
Uncertainties on every data point calculated from photon statistics are
rigorously propagated through every stage of the data reduction. 

   Finally, a single spectrum of PG\,0940+068 was obtained with the
cross-dispersed echelle spectrograph UCLES on the AAT.  Details of the
reduction of these spectra are similar to those described in Gatti et~al. 
(1997). 

 Details of all these instruments, the dates of all the observing runs and the
resolution and sampling obtained are given in Table~\ref{ObsTable}.

\section{PG\,0940+068}
 This star was first noted as a potential binary by Bragaglia et~al. (1990)
and later confirmed as such by Saffer et~al. (1998). Saffer et~al. also noted
the presence of the He\,I~6678 absorption line in their spectra which
identifies this star as an sdB star rather than a white dwarf.

 To measure the radial velocities we used least squares fitting of a model
line profile. This model line profile is the summation of three Gaussian
profiles with different widths and depths but with a common central position
which varies between spectra. The resolution of each instrument is accounted
for in the fitting process. Only data within 2000\,km/s of the H$\alpha$ line
is included it the fitting process. We first normalise the spectra using a
linear fit to the continuum either side of the H$\alpha$ line. We used a
least-squares fit to a single spectrum to establish the shape of an initial
model line profile. A least squares fit of this profile to each spectrum in
which the position of the line is the only free parameter yields an initial set
of radial velocities. The periodogram of these velocities shows the best
period to be near 8.33d. Other peaks in the periodogram are much less
significant. We used a sine fit to the radial velocities to fix the position
of the H$\alpha$ line in each spectrum in a simultaneous least squares fit to
all the spectra to establish an optimum model line profile. We then
re-measured the radial velocities of each spectrum using this optimum model
line profile. The quality of the model line fits to the spectra estimated
by-eye and from the chi-square statistic were generally good. These are the
radial velocities given in Table~\ref{0940RV}. 

\begin{figure*} 
\caption{\label{0940Fig} Measured radial velocities for PG\,0940+068 and the
 circular orbit fit.}
\psfig{file=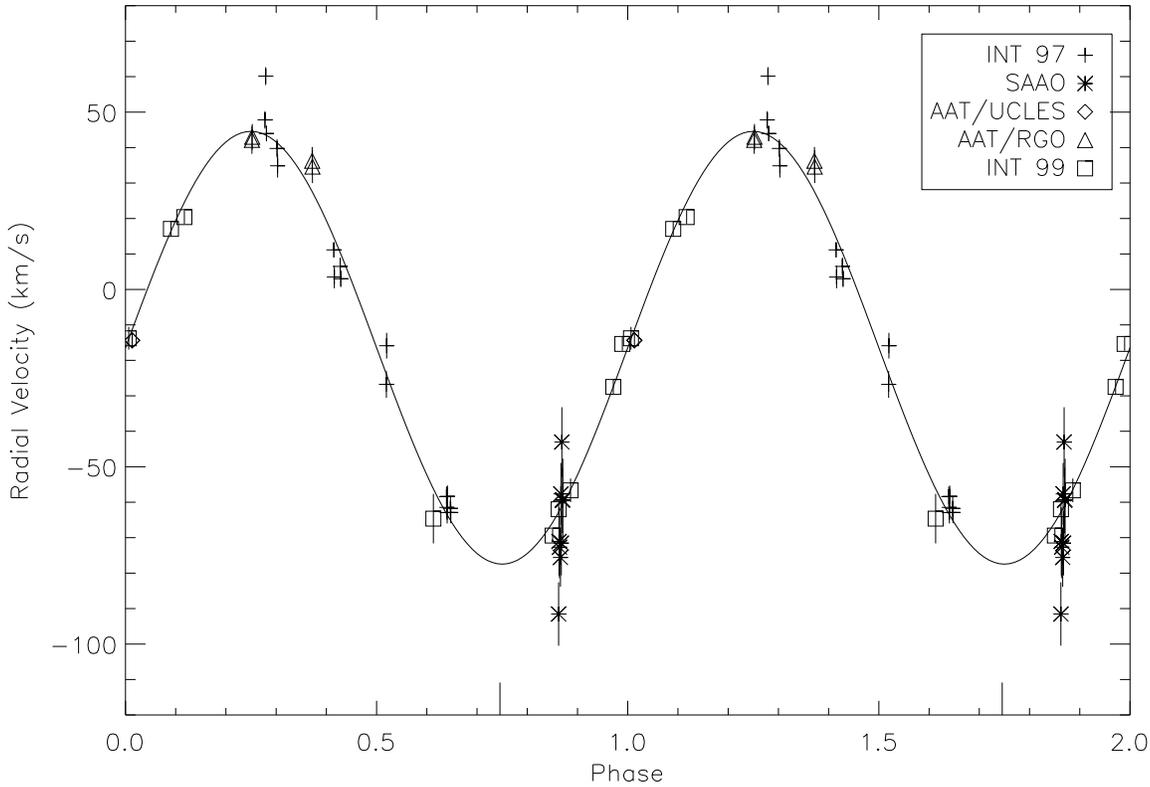,width=0.9\textwidth} 
\end{figure*}

 The parameters of a circular orbit fit to the radial velocities are
given in Table~\ref{FitTable}. The value of chi-squared is rather high.
This is, perhaps, not surprising given the diverse sources of spectra on
which this orbit is based, but might also suggest that there are
additional sources of uncertainty in our measured radial velocities,
e.g., image motion within the slit, instrument flexure and telluric
absorption lines.  We have adjusted the uncertainties in the measured
parameters accordingly, i.e., multiplied them by $\sqrt{141.1/(41-4)}$.
The measured radial velocities and the circular orbit fit are shown in
Fig.~\ref{0940Fig}.

 We also established the parameters of the orbit by fitting all the spectra
simultaneously using a least squares fit in which the position of the
absorption line is determined by a circular orbit. The parameters of the orbit
and the model line profile are all free parameters in the fitting process.
This gives essentially the same result as fitting the individual radial 
velocities. We also  subtracted the model profiles from the observed spectra
to look for spectral features due to a companion star, but without success.

\begin{table*}
\caption{\label{0940RV} Radial velocity measurements for PG\,0940+068}
\begin{tabular}{rrcrrcrrcrr}
\multicolumn{1}{l}{HJD} & Velocity &~&
\multicolumn{1}{l}{HJD} & Velocity &~&
\multicolumn{1}{l}{HJD} & Velocity &~&
\multicolumn{1}{l}{HJD} & Velocity \\
-2450000 & (km\,s$^{-1}$) &~&
-2450000 & (km\,s$^{-1}$) &~&
-2450000 & (km\,s$^{-1}$) &~&
-2450000 & (km\,s$^{-1}$) \\
852.4349&  47.6$\pm$2.5&~&855.4455& -58.2$\pm$ 2.9&~& 882.3250& -75.5$\pm$ 7.7&~& 969.8409&  34.6$\pm$ 4.1 \\
852.4463&  60.0$\pm$2.6&~&855.4511& -61.8$\pm$ 3.1&~& 882.3323& -56.7$\pm$ 8.1&~& 1238.4055&-64.2$\pm$ 7.1 \\
852.4589&  43.8$\pm$2.2&~&855.4561& -62.9$\pm$ 3.1&~& 882.3397& -71.4$\pm$ 8.4&~& 1240.3832&-69.3$\pm$ 2.0 \\
852.6349&  39.7$\pm$2.6&~&855.4611& -58.3$\pm$ 3.1&~& 882.3484& -49.2$\pm$ 9.2&~& 1240.4840&-61.9$\pm$ 1.8 \\
852.6434&  34.9$\pm$3.4&~&855.5102& -61.6$\pm$ 3.2&~& 882.3567& -59.2$\pm$ 8.6&~& 1240.6804&-56.5$\pm$ 3.4 \\
853.5760&  11.2$\pm$2.2&~&855.5151& -62.7$\pm$ 3.1&~& 882.3640& -57.8$\pm$10.8&~& 1241.3896&-27.4$\pm$ 2.1 \\
853.5847&   3.5$\pm$3.2&~&881.3208&-131.8$\pm$18.1&~& 941.8613& -13.4$\pm$ 1.6&~& 1241.5396&-15.4$\pm$ 1.9 \\
853.6816&   6.4$\pm$2.5&~&882.2942& -92.5$\pm$ 8.3&~& 968.8381&  41.9$\pm$ 3.4&~& 1241.6821&-13.8$\pm$ 3.2 \\
853.6912&   3.1$\pm$2.4&~&882.3016& -71.5$\pm$ 9.0&~& 968.8419&  43.1$\pm$ 3.2&~& 1242.3831& 17.0$\pm$ 2.1 \\
854.4479& -26.4$\pm$3.8&~&882.3089& -72.7$\pm$ 8.1&~& 969.8371&  36.4$\pm$ 3.4&~& 1242.6059& 20.5$\pm$ 1.9 \\
854.4519& -15.7$\pm$3.7 \\
\end{tabular}
\end{table*}

\section{PG\,1247+554 (GD\,319)}
  This star was first noted as a potential binary by Saffer et~al. (1998), who
also noted the strong He\,I~6678 line which suggests that this star is a blue
horizontal branch  (BHB) star or low-gravity sdB star. There is a  K star 
companion 2.7arcsec away but this was shown to be unrelated to PG\,1247+554 by
McAlister et~al (1996), i.e., it is an optical double star.

 We acquired 11 spectra of this star and its nearby companion star over 4
nights of March 1999 covering the H$\alpha$ line and the He\,I~6676 line, 
4 spectra over two nights of June 1997 covering the H$\alpha$ line only
and two spectra covering the H$\alpha$ line and the He\,I 6676 line on a
night of July 1999.

We same procedure described above for PG\,0940+068 to establishing a model
line profile and measure the radial velocities from the H$\alpha$ line. These
are given in column 2 of Table~\ref{1247RV}.  We also measured radial
velocities from the HeI\,6678 line in our 1999 spectra using a similar
procedure but using only two Gaussian profiles to model the line. The
velocities are given in column 3  of Table~\ref{1247RV}. 

\begin{figure*} 
\caption{\label{1247Fig} Measured radial velocities for PG\,1247+554 and the
 circular orbit fit.}
\psfig{file=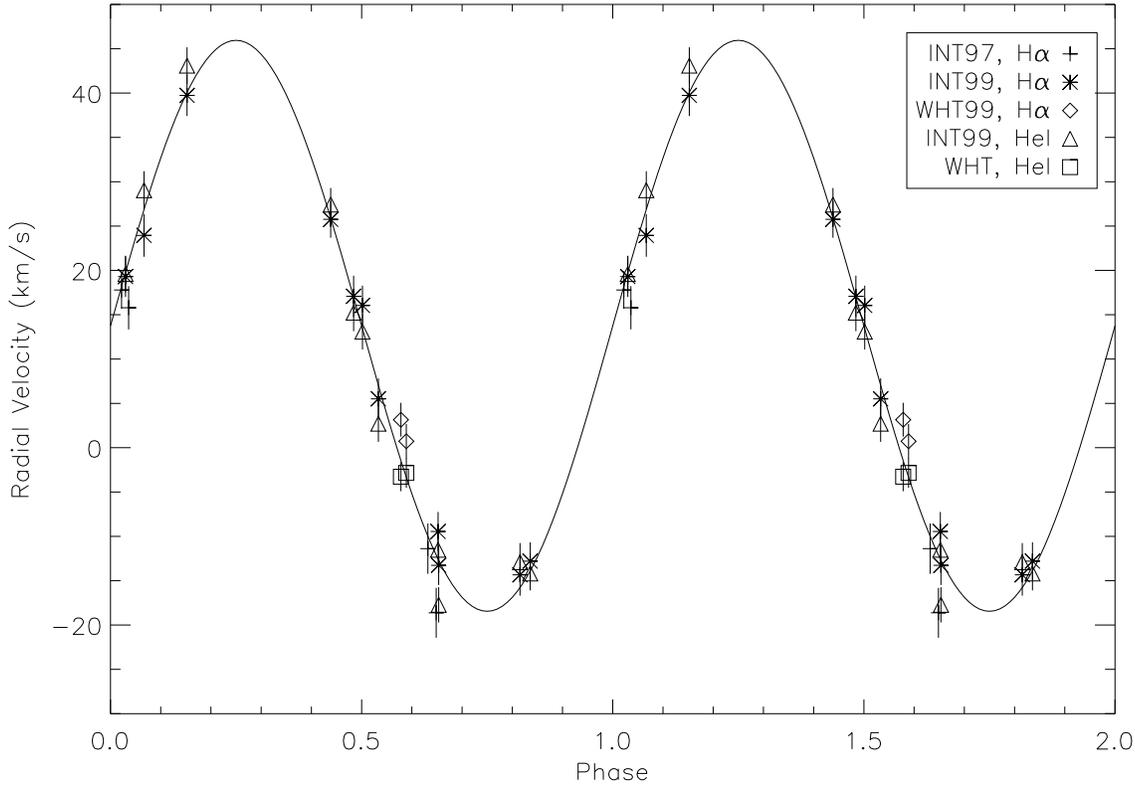,width=0.9\textwidth} 
\end{figure*}

 The parameters of a circular orbit fit to the radial velocities
measured from both spectral lines are given in Table~\ref{FitTable} and
the fit is shown in Fig.~\ref{1247Fig}. The value of chi-squared is,
again, rather high so we have adjusted the uncertainties in the
measured parameters accordingly, i.e., multiplied them by
$\sqrt{45.3/(30-4)}$.  A simultaneous fit to all the spectra to
determine the orbit gives essentially the same results and there is no
trace of a companion star in the residual spectra.

\begin{table}
\caption{\label{FitTable} Circular orbit fits. The systemic velocity is
denoted by $\gamma$ and the semi-amplitude by K.}
\begin{tabular}{lrr}
Parameter&\multicolumn{1}{l}{PG\,0940+068}&\multicolumn{1}{l}{PG\,1247+554} \\
HJD(T$_0$)-2450000      & 1000.066 $\pm$ 0.027 & 1240.855 $\pm$ 0.002 \\
Period(d)               & 8.330 $\pm$ 0.003   &        0.602740 \\
                        &                     & $\pm$  0.000006 \\
$\gamma$ (km\,s$^{-1}$) & -16.7 $\pm$ 1.4     & 13.8 $\pm$ 0.6 \\
K (km\,s$^{-1}$)        &  61.2 $\pm$ 1.4     & 32.2$\pm$ 1.0 \\
Mass function(\Msolar)  & 0.20 $\pm$ 0.01     & 0.0021 $\pm$ 0.0002 \\
$\chi^2$                & 141.1               & 45.3 \\
n                       & 41                  & 30  \\
\end{tabular}
\end{table}

\begin{table}
\caption{\label{1247RV} Radial velocity measurements for PG\,1247+554}
\begin{tabular}{lrrr}
HJD & \multicolumn{1}{l}{H$\alpha$} & \multicolumn{1}{l}{HeI\,6678} \\
-2450000 & \multicolumn{1}{l}{(km\,s$^{-1}$} \\
 622.4567 &  17.7  $\pm$ 2.1  & --- \\
 622.4652 &  15.7  $\pm$ 2.4  & --- \\
 623.4270 & -11.3  $\pm$ 2.8  & --- \\
 623.4370 & -18.6  $\pm$ 2.8  & --- \\
1240.5436 &  17.0  $\pm$ 2.3  &  15.2 $\pm$ 2.0  \\
1240.5541 &  16.0  $\pm$ 2.2  &  13.0 $\pm$ 2.0  \\
1240.6456 & -13.2  $\pm$ 2.2  & -17.7 $\pm$ 1.9  \\
1240.7433 & -14.3  $\pm$ 2.3  & -12.9 $\pm$ 2.1  \\
1241.4975 &  23.9  $\pm$ 2.4  &  29.0 $\pm$ 2.1  \\
1241.7215 &  25.7  $\pm$ 2.0  &  27.4 $\pm$ 1.8  \\
1241.7788 &   5.5  $\pm$ 2.3  &   2.7 $\pm$ 2.0  \\
1242.5637 & -12.7  $\pm$ 2.1  & -14.1 $\pm$ 1.9  \\
1242.6808 &  19.2  $\pm$ 2.2  &  19.5 $\pm$ 2.0  \\
1242.7546 &  39.7  $\pm$ 2.3  &  43.0 $\pm$ 2.0  \\
1243.6586 &  -9.4  $\pm$ 2.1  & -11.5 $\pm$ 1.9  \\
1368.3812 &   3.1  $\pm$ 1.9  &  -3.2 $\pm$ 1.6  \\
1368.3877 &   0.7  $\pm$ 1.9  &  -2.8 $\pm$ 1.6  \\
\end{tabular}
\end{table}

\section{Discussion}
 If we assume a canonical mass of 0.5\Msolar\ for the visible components of
these binaries (Saffer et~al. 1994), we can calculate a minimum mass of
$0.63\pm0.02$\Msolar\ for the unseen companions  to PG\,0940+068 and
$0.036\pm0.003$\Msolar\ for PG\,1247+554. If the unseen companion  to
PG\,0940+068 were a main-sequence star it would have an absolute magnitude,
M$_{\rm V} = 8.3\pm$0.9 (Henry \& McCarthy 1993), where the uncertainty
reflects the error in the mass and the scatter in the mass\,--\,luminosity
relation. On average, the absolute magnitude of an sdB star is M$_{\rm V}
=4.5\pm0.8$, so a main-sequence companion is likely to be contribute at least
1percent of the light at V. While this would not have been seen in our
spectra, it should allow the nature of the companion to be established through
high signal-to-noise spectroscopy, perhaps around the Ca\,II IR triplet as in
Jeffery \& Pollacco (1998). 

\section{Conclusion} 
 We have measured the orbital periods for two sdB binary stars and calculated
the minimum mass of the companion star. The orbital period of  PG\,0940+068 is
8.33d and the minimum mass of the companion is $0.63\pm0.02$\Msolar. The
orbital period of PG\,1247+554 is 0.598655d and the minimum mass of the
companion is $0.087\pm0.003$\Msolar. If the companion to PG\,0940+068 is a
main sequence star, it will contribute 1percent of the light at V, which may
make it directly detectable.

\section*{Acknowledgements}
 PFLM was supported by a PPARC post-doctoral grant. CM was supported by a
PPARC post-graduate studentship. The William Herschel Telescope and the Isaac
Newton Telescope are operated on the island of La Palma by the Isaac Newton
Group in the Spanish Observatorio del Roque de los Muchachos of the Instituto
de Astrofisica de Canarias.

\label{lastpage}

\end{document}